\documentclass[twocolumn,final,aps,prb,showpacs,superscriptaddress,amsmath,amssymb,amsfonts,floatfix]{revtex4-2}
\usepackage{epsfig}
\usepackage[dvipsnames,usenames]{color}
\usepackage{color}
\usepackage{amsmath}
\usepackage{physics}
\usepackage{comment}
\usepackage{array}
\usepackage{multirow}
\usepackage{tabularx}

\usepackage{ulem}
\usepackage{placeins}
\usepackage[colorlinks=true,bookmarks=true, pdfborder={0 0 0},linkcolor=blue,urlcolor=blue,    breaklinks=true,bookmarksnumbered=true]{hyperref}
\usepackage{orcidlink}

\newcommand{\jgw}[1]{{\color{black}#1}}

\emergencystretch=\maxdimen
\begin{document}
\title{Pressure induced redistribution of oxygen hole states in La$_4$Ni$_3$O$_{10}$}
\author{Guiwen Jiang \orcidlink{0009-0006-8807-4849}}

\affiliation{Institute for Quantum Science, School of Physical Science and Technology, Soochow University, Suzhou 215006, China}

\author{Liang Si}
\affiliation{School of Physics, Northwest University, Xi'an 710127, China}
\affiliation{Institute of Solid State Physics, TU Wien, 1040 Vienna, Austria}

\author{George A. Sawatzky}
\affiliation{Department of Physics and Astronomy, University of British Columbia, Vancouver BC, Canada V6T 1Z1}
\affiliation{Stewart Blusson Quantum Matter Institute, University of British Columbia, Vancouver BC, Canada V6T 1Z4}

\author{Mi Jiang \orcidlink{0000-0002-9500-202X}}
\email[]{jiangmi@suda.edu.cn}
\affiliation{Institute for Quantum Science, School of Physical Science and Technology, Soochow University, Suzhou 215006, China}
\affiliation{State Key Laboratory of Surface Physics and Department of Physics, Fudan University, Shanghai 200433, P. R. China}

\begin{abstract}
Using density functional calculations and multi-orbital, multi-atom cluster exact diagonalization that includes local exchange and Coulomb interactions, we explored the local low-energy electronic states of trilayer La$_4$Ni$_3$O$_{10}$ via a minimal Ni$_3$O$_{14}$ cluster. 
We find that, at ambient pressure, starting with all three Ni being nominally 2+ valence, one of the two extra holes is localized in the central NiO$_2$ layer forming a Zhang-Rice singlet (ZRS) with $d_{x^2-y^2}$ orbital. The other hole mainly occupies the antibonding combination of the two interplane apical O $p_z$ orbitals and thereby hybridizes with an out-of-plane three-spin-polaron (3SP) formed by the $d_{z^2}$ orbitals of three NiO$_2$ layers. 
\jgw{At high pressure, however, the two extra holes are concentrated on one of two outer layers and the inner layer separately forming the ZRS with $d_{x^2-y^2}$ orbitals. 
We highlight the similarities between the bilayer La$_3$Ni$_2$O$_7$ and trilayer La$_4$Ni$_3$O$_{10}$ via speculated possible charge and spin configurations as well as the in-plane 3SP on two neighboring clusters suggested by our isolated cluster results.
We thereby propose that the hole transfer from apical to in-plane oxygen orbitals of outer layer generates in-plane 3SP-like quasiparticles that act as mobile carriers coupled by interlayer superexchange; while the interplane 3SP-like states may provide the pairing glue. Since the low-pressure phase lacks freely propagating in-plane quasiparticles, this scenario naturally favors SC in the high-pressure phase.}
\end{abstract}

\maketitle

\section{Introduction}
Recently, superconductivity (SC) has been discovered in both bilayer and trilayer Ruddlesden-Popper (RP) nickelate single crystals under pressure, with maximum transition temperatures $T_c$ reaching about 80 K and 30 K, respectively~\cite{sun2023signatures,zhu2024superconductivity,zhang2025superconductivity}. Compared to La$_3$Ni$_2$O$_7$, La$_4$Ni$_3$O$_{10}$ exhibits better sample quality and greater stability, although its superconducting transition temperature is significantly lower than that of La$_3$Ni$_2$O$_7$, possibly due to weaker electronic correlations in La$_4$Ni$_3$O$_{10}$~\cite{xu2025origin,li2025direct}. With increasing pressure, the density-wave order in La$_4$Ni$_3$O$_{10}$ is suppressed, accompanied by a structural transition from the $P2_1/a$ phase to the $I4/mmm$ phase, accompanied by the emergence of SC~\cite{zhu2024superconductivity,zhang2025superconductivity}. The interplay between the crystal structure, density-wave order, and SC is highly intricate. Several studies have shown that multiple other phases or crossovers occur between the low- and high pressure phases~\cite{zhu2024superconductivity,zhang2025superconductivity, xu2025collapse}, indicating the close proximity of various electronic and magnetic structures.

Numerous experimental studies have demonstrated a close correlation between SC and density-wave order~\cite{xu2025collapse,xu2025origin,li2025direct,shi2025pressure,zhao2025pressure,jia2026lattice}. In La$_4$Ni$_3$O$_{10}$ microcrystals synthesized in the $I4/mmm$ phase at ambient pressure, no signature of density-wave formation or SC has been observed even under pressures up to 160~GPa~\cite{shi2025absence}. This observation suggests that the density-wave order, rather than the structural transition itself, plays a crucial role in the emergence of SC. Further investigation revealed that the spin-density-wave (SDW) and charge-density-wave (CDW) components remain coupled as pressure increases~\cite{li2025direct,xu2025collapse,nrqn-m22c,wang2026unconventional,guan2026electroniclayer}. The SDW transition exhibits a first-order-like character, and its transition temperature coincides with that of the CDW~\cite{nrqn-m22c}. In addition, two independent $\mu$SR studies have reported an additional SDW order emerging at a lower temperature of $T_\mathrm{SDW}\approx80$~K in  single-crystalline and polycrystalline samples of La$_4$Ni$_3$O$_{10}$~\cite{nrqn-m22c,cao2025complex}.

Exploring electronic structures is essential to gain insights into the superconducting mechanism of La$_4$Ni$_3$O$_{10}$. Angle-resolved photoemission spectroscopy (ARPES) measurements and DFT calculations reveal that, in addition to $d_{x^2-y^2}$ orbital, La$_4$Ni$_3$O$_{10}$ also exhibits a significant $d_{3z^2-r^2}$ orbital character~\cite{li2017fermiology,electronicorigindensitywave}. However, there remains considerable debate regarding which orbital plays the dominant role in SC. In particular, recent orbital-selective resonant inelastic X-ray scattering (RIXS) measurements indicate that the $d_{x^2-y^2}$ orbital dominates the low-energy charge excitations and exhibits more itinerant behavior~\cite{zhang2025distinct}. They also observed a $\sim 0.1$~eV bimagnon excitation using RIXS and Raman spectroscopy, leading to an interlayer superexchange interaction $J_z$ of $\sim 50$~meV~\cite{zhang2025distinct}. \jgw{Other studies based on linear spin wave theory reported an interlayer superexchange $J_z$ as small as $\sim 22$~meV, which is much smaller than that in La$_3$Ni$_2$O$_7$~\cite{Dissecting,collectivespin}.} 

Theoretical studies based on density functional theory combined with dynamical mean-field theory (DFT+DMFT) have extensively investigated the electronic correlations and the role of Hund's coupling in La$_4$Ni$_3$O$_{10}$~\cite{huo2025electronic,tian2024effective}. Another DFT+DMFT simulation has shown that the inner/outer NiO$_2$ layer exhibits weak/strong Hund correlation~\cite{wang2024non}. Several theoretical studies based on random phase approximation (RPA) have suggested that this system  predominantly supports the $s^{\pm}$-wave pairing~\cite{zhang2024prediction,zhang2024s,swavela4ni310}, consistent with the conclusion obtained from the minimal tight-binding model~\cite{yang2024effective}. In addition, DFT+U calculations under pressure have revealed an upward shift of the Ni-$d_{3z^2-r^2}$ bonding band and a reconstruction of the electronic states, while RPA and multi-orbital analysis emphasized the role of $d_{x^2-y^2}$ magnetism and charge-transfer character~\cite{luo2024trilayer}.

Compared with La$_3$Ni$_2$O$_7$, La$_4$Ni$_3$O$_{10}$ possesses additional orbital and layer degrees of freedom, making many numerical approaches challenging to apply.
Although exact diagonalization (ED) is constrained to small clusters, thus introducing finite-size effects, its significance lies in producing numerically exact solutions. It avoids the sign problem that limits methods such as determinant quantum Monte Carlo (DQMC) and allows for the direct computation of zero-temperature ground and excited states. This precision, combined with the relative ease of incorporating complex interactions such as those in multi-orbital systems with Hund's coupling, establishes ED as a critical benchmark for theories of strong correlation. The remainder of this paper is organized as follows. In Section \ref{Model and Method}, we introduce the DFT calculation and the multi-orbital cluster model. Section \ref{sec:results} presents the numerical results and analysis. Finally, in Section \ref{sec:Summary} we give a summary and a discussion.

\section{Model and Methodology}
\label{Model and Method}
\subsection{DFT calculations and Wannier projections}
The Density-functional theory~\cite{PhysRev.136.B864,PhysRev.140.A1133} level structural relaxations and band structure calculations were performed using the \textsc{VASP}~\cite{kresse1996efficiency,PhysRevB.54.11169} and \textsc{Wien2K}~\cite{blaha2001wien2k,schwarz2002electronic} codes with the Perdew-Burke-Ernzerhof version of the generalized gradient approximation (GGA-PBE)~\cite{PhysRevLett.77.3865} and a dense mesh of more than 2000 $k$-points for the unit cell of La$_4$Ni$_3$O$_{10}$. The parameters listed in Table I are determined from the corresponding DFT Bloch bands and Maximally localized Wannier functions~\cite{PhysRev.52.191,RevModPhys.84.1419} projections using  \textsc{Wannier90}~\cite{mostofi2008wannier90}  and \textsc{WIEN2WANNIER}~\cite{kunevs2010wien2wannier} interface. The most crucial parameters in all three models are the hybridization tds between Ni-3$d_{x^2-y^2}$, Ni-3$_{dz^2}$, in-plane and apical O-2$p$ orbitals, and these hopping parameters between Ni-3$d$ orbitals, Ni-3$d$/O-2$p$ to O-2$p$ were obtained via the Wannier projections for models of Ni-3$d$ subspace and Ni-3$d$ and O-2$p$ substate, respectively. 

The hole hopping parameters for the $P2_1/a$ phase at 0 GPa and the $I4/mmm$ phase at 15.3 GPa are listed in Table~\ref{parameter}.

\begin{table*}[t!]
\centering
\caption{On-site energies $\epsilon$, Racah parameters $A$, $B$, and $C$, together with hopping integrals $T^{pd}_{mn}$ extracted from DFT calculations. Here $m \in \{d_{x^2}, d_{z^2}\}$, where $d_{x^2}\equiv d_{x^2-y^2}$ and $d_{z^2}\equiv d_{3z^2-r^2}=d_{2z^2-(x^2-y^2)}$, while $n\in\{p_x, p_y\}$. The corresponding orbitals involved in $T^{pd}_{mn}$ are nearest neighbors. \jgw{Throughout this work, $p$ denotes the in-plane $p_x$ or $p_y$ orbital whose lobe points toward the Ni ion, whereas $O$ denotes the $p_z$ orbital of the interlayer apical oxygen. }}
\setlength{\tabcolsep}{19pt}
\begin{tabular}{*{8}{c}}
\hline
$T^{pd}_{x^2-y^2, n}$ & $T^{pd}_{z^2, n}$ & $e_{dx^2}$ & $A$ & $B$ & $C$ & $U_{OO}$ & $U_{pp}$ 
 \\[0.5ex]
\hline
$t_{pd}$ & $t_{pd}/\sqrt{3}$ & 0 & 6 & 0.15 & 0.58 & 4.0 & 4.0 \\[0.5ex]
\hline\hline
0~GPa & $e_{dz^2}$ & $e_{p}$ & $e_{O}$ & $t_{pd}$ & $t_{pp}$ & $t_{do}$ & $t_{po}$ \\[0.5ex] \hline
Inner & 0.02 & 2.875 & 2.76 & 1.285 & 0.435 & 1.210 & 0.435 \\[0.5ex]
Outer & 0.17 & 3.093 & \text{/} & 1.187 & 0.478 & 1.266 & 0.403 \\[0.5ex]
\hline

15.3~GPa & $e_{dz^2}$ & $e_{p}$ & $e_{O}$ & $t_{pd}$ & $t_{pp}$ & $t_{do}$ & $t_{po}$ \\[0.5ex] \hline
Inner & -0.076 & 3.364 & 3.047 & 1.503 & 0.525 & 1.625 & 0.489 \\[0.5ex]
Outer & 0.116 & 3.519 & \text{/} & 1.498 & 0.551 & 1.666 & 0.486 \\[0.5ex] \hline
\end{tabular}
\label{parameter}
\end{table*}

\subsection{Multi-orbital cluster Model}

\begin{figure}[t!]
\centering
\psfig{figure=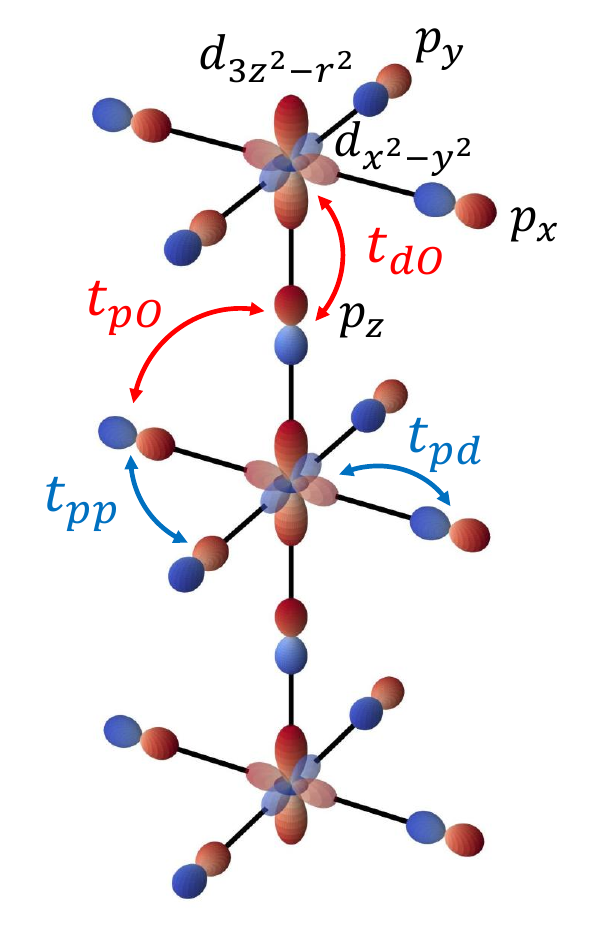, width=.45\textwidth, clip}
\caption{Geometry of the $\mathrm{Ni_{3}O_{14}}$ cluster used in the calculations. \jgw{The figure shows the Ni-$e_g$ and oxygen orbitals retained in the configuration basis, together with various hoppings included in the model Hamiltonian.}}
\label{geometry}
\end{figure}

Here we extend our previous study~\cite{jiang2025intertwined} of the bilayer Ni$_2$O$_9$ cluster to a trilayer Ni$_3$O$_{14}$ cluster comprising three NiO$_2$ planes separated by apical oxygen atoms as depicted in Fig.~\ref{geometry}. As shown before~\cite{jiang2025intertwined}, the bilayer cluster has indicated that 4 of the total 5 holes occupy the two Ni $3d^8$ sites and the remaining hole mainly occupies the in-plane O $p_x$ orbitals. Because of the local nature of this minimal cluster, we focus on the intrinsic local physics induced by the interplay of the hybridization between various orbitals and interactions among these orbitals.

Assuming a fixed $-2$ valence for oxygen, the total formal valence of the three Ni ions in La$_4$Ni$_3$O$_{10}$ sums up to $+8$, producing an average Ni valence of $+2.67$, commonly described as a mixture of Ni$^{2+}$ and Ni$^{3+}$. However, compared to the bilayer La$_3$Ni$_2$O$_7$, the hole distribution in this trilayer system is considerably more complex. Presumably, in the hole language adopted in our previous work~\cite{jiang2025intertwined}, 6 of 8 holes would occupy the three Ni $3d^8$ sites so that the key question is where the remaining two holes reside among different layers and orbitals, which is largely affected by the pressure induced change of various hybridization. 

We employ the exact diagonalization calculation of the Hamiltonian
\begin{align} 
\mathcal{H}= &\hat{E}_d+\hat{E}_p+\hat{E}_O+ \hat{T}_{pp}+\hat{T}_{dp}+\hat{T}_{dO}+\hat{T}_{pO} \notag \\
&+
\hat{U}_{dd}+\hat{U}_{pp}+\hat{U}_{OO}
\end{align}
\noindent where $\hat{E}_d$, $\hat{E}_p$ and $\hat{E}_O$ are the on-site energies for the orbitals of Ni, the oxygen ligand $2p$, and the interlayer apical oxygen. As usual, the site energy of Ni-$3d_{x^2-y^2}$ is set to zero as a reference. 
\jgw{Here, $\hat{T}_{pp}$, $\hat{T}_{dp}$, $\hat{T}_{dO}$ and $\hat{T}_{pO}$ represent the hopping integrals between the retained orbitals. The symbol O denotes the $p_z$ orbital of the interlayer apical oxygen, whereas $p$ refers to the in-plane oxygen orbitals.} $\hat{U}_{dd}$, $\hat{U}_{pp}$, and $\hat{U}_{OO}$ describe their on-site Coulomb interactions, respectively~\cite{eskes1990cluster}. Specifically, $\hat{U}_{dd}$ is the Coulomb and exchange integrals corresponding to the symmetry $D_{4h}$ in terms of Racah parameters $A, B, C$, which are connected to the Slater integrals $F^0$, $F^2$, and $F^4$ for the transition metal $3d^8$ multiplets as~\cite{Sugano1970} 
\begin{align}
A &= F^0-\frac{49}{441}F^4 \notag \\
B &= \frac{1}{49}F^2 - \frac{5}{441}F^4 \notag \\
C &= \frac{35}{441}F^4
\end{align}
The parameter $A$ is analogous to the Hubbard $U$ commonly used in nickelates, while $B$ and $C$, derived from atomic physics, quantify the energy differences for the two holes in the Ni $3d$ shell, including their dependence on the relative spin and orbital orientation. 
The conventional Racah parameters for Ni, with $A = 6.0$ eV, $B = 0.15$ eV, and $C = 0.58$ eV are adopted similar to our previous work~\cite{jiang2020critical}. 
The DFT calculations indicate that low-energy physics is dictated by the Ni $e_g$ orbitals. Hence, our Hilbert space is restricted to these two Ni $e_g$ orbitals and the O-$2p$ orbitals that hybridize with them through $\sigma$ bonds (as shown in Fig.~\ref{geometry}).

Furthermore, the on-site Hubbard interaction for two holes in the same oxygen atom has been measured by Auger spectroscopy in oxides, yielding a value of 4.6~eV in Cu$_2$O~\cite{ghijsen1988electronic}. Given the relatively smaller ionic radius of Ni compared to Cu, we adopt an estimated value of 4.0~eV for both the apical and in-plane oxygen sites. 

To denote the multi-hole configuration, \jgw{throughout this work we use a notation of the form $d^8$-$d^8L$-$O$-$d^8$ to specify the configuration. Here, $d^8$ denotes a Ni ion with eight electrons in the $3d$ shell. $L$ represents a ligand hole on the surrounding oxygen orbitals, precisely a linear combination of the nearest-neighbor oxygen ligand orbitals adapted to the symmetry of a specific Ni $3d$ orbital. For example, the $x^2-y^2$ and $3z^2-r^2$ symmetric ligand combinations hybridize with the Ni $3d_{x^2-y^2}$ and $3d_{3z^2-r^2}$ orbitals, respectively. In addition, $O$ denotes a hole in the $p_z$ orbital of interlayer apical oxygen. Therefore, in the above example, the three Ni layers are in the $d^8$, $d^8L$, and $d^8$ configurations separately. 

It should be noted that two mirrored configurations are combined when evaluating the weights. For example, the reported weight of the $d^8$-$d^8L$-$O$-$d^8$ configuration includes the contribution from its mirror counterpart $d^8$-$O$-$d^8L$-$d^8$.
In addition, we perform a single-particle bonding and antibonding transformation on their apical $p_z$ orbitals
\begin{align}
O_{p_z}(B) &= \frac{1}{\sqrt{2}}(p_{z_1} + p_{z_2}) \notag \\
O_{p_z}(A) &= \frac{1}{\sqrt{2}}(p_{z_1} - p_{z_2})
\end{align}
We find that the GS weight associated with the antibonding orbital is always dominant, while only a small fraction resides in the bonding orbital at 0~GPa. For example, the bonding-antibonding splitting energy is found to be 0.96~eV at 0~GPa.} 
With increasing pressure, a phase transition occurs, and the dominant configuration becomes $d^8$-$d^8L$-$d^8L$.

Besides, for $d^8$ states involving {$d_{z^2}d_{x^2}$}, curly braces \{.\} are used to indicate spin pairs coupled into triplet states. Here, the shorthands $x = d_{x^2} \equiv d_{x^2-y^2}$ and $z = d_{z^2} \equiv d_{3z^2-r^2}$ are adopted. The total spin $S$, which is a good quantum number for our cluster model, characterizes the spin quantum number of the various 7-, 8-, and 9-hole states.

\subsection{Energy level diagram}

Before solving the Ni$_3$O$_{14}$ cluster, we first consider the atomic limit to determine the hole distribution and the corresponding energies. Similarly to La$_3$Ni$_2$O$_7$, the nominal ground-state configuration of the Ni ions in La$_4$Ni$_3$O$_{10}$ is $d^8$, which is taken as the energy reference. 

We start from the definition of the Hubbard interaction 
\begin{align}
U=E(d^{n+1})+E(d^{n-1})-2E(d^n)
\label{eq_U}
\end{align} 
and the onsite energy of the $d$ orbital in the hole language 
\begin{align}
\epsilon_d = E(d^9)-E(d^{10})
\label{eq_onsite}
\end{align}
Using Eqs.~(\ref{eq_U}) and (\ref{eq_onsite}), we obtain the following
\begin{align}
E(d^{10})&=0 \notag \\
E(d^9)&=\epsilon_d \notag \\
E(d^8)&=2\epsilon_d + U \notag \\
E(d^7)&=3\epsilon_d + 3U\notag \\
E(d^6)&=4\epsilon_d+6U
\label{eq_Edn}
\end{align}
We assume that the electron addition and removal energies from the $d^8$ configuration are symmetric, satisfying
\begin{align}
E(d^9) - E(d^8) = E(d^7) - E(d^8) = \frac{U}{2} 
\label{eq_d9-d8}
\end{align}
Substituting Eqs.~(\ref{eq_Edn}) into Eq.~(\ref{eq_d9-d8}) gives $\epsilon_d = - \frac{3}{2}U$, which when substituted again into Eqs.~(\ref{eq_Edn}), yields
\begin{align}
E(d^8)&=-2U \notag \\
E(d^7)&=E(d^9) =-\frac{3}{2}U \notag \\
E(d^6)&=E(d^{10})=0
\end{align}
Taking the $d^8$ configuration as the energy reference state and shifting the overall energy by $2U$, we have \begin{align}
E(d^8)&=0 \notag \\
E(d^7)&=E(d^9) =U/2 \notag \\
E(d^6)&=E(d^{10})=2U
\end{align} 

\begin{figure}[t!]
\psfig{figure=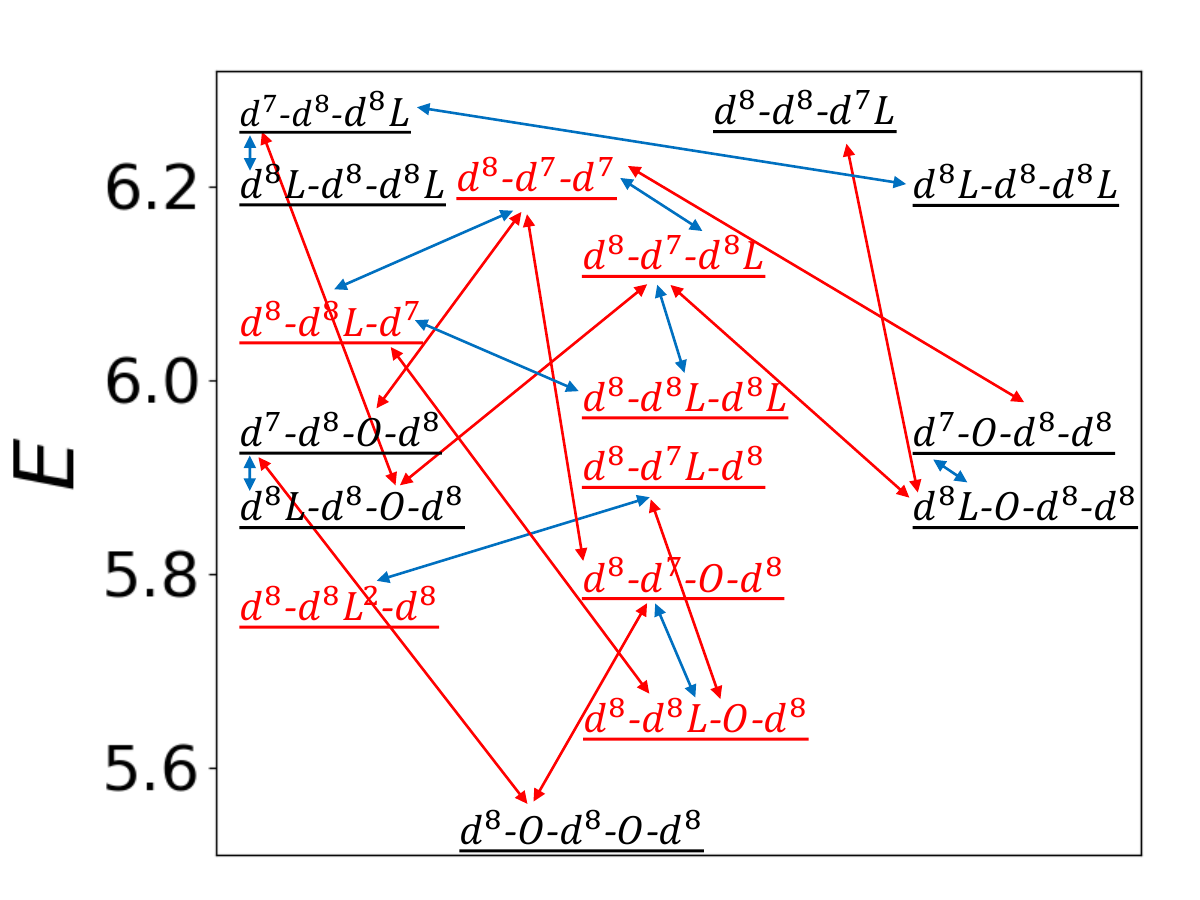, width=0.45\textwidth, clip}
\caption{Energy level diagram of  8-hole cluster in the absence of hybridization and correlations apart from the Ni $U$ with the configurations of even higher energies omitted. \jgw{Here, $d^n$ denotes a Ni ion with n electrons in the $3d$ shell. The symbol $L$ represents a ligand hole residing in a linear combination of the nearest-neighbor oxygen ligand orbitals adapted to the symmetry of a specific Ni $3d$ orbital and $O$ denotes a hole on the interlayer apical oxygen $p_z$ orbital.} The red (blue) arrow indicates $t_{dO}$ ($t_{pd}$) hybridization matrix elements. The energy of $d^8$ is chosen as the zero energy reference. These configurations are highly degenerate in the atomic limit. Once the various hopping integrals are switched on, the degeneracy is lifted, and the configurations that carry dominant weights in the ground state of the 8-hole system are highlighted in red.}
\label{diagram_8hole}
\end{figure}

With these term energies, we can have a full understanding of all significant low-energy configurations of multi-hole systems. To this end, Fig.~\ref{diagram_8hole} illustrates the energy level diagram in the atomic limit. Here, with the Racah parameter A denoting the interaction strength akin to the Hubbard $U$ in a single-orbital, we have $U \approx A=6.0$~eV. As shown in Fig.~\ref{diagram_8hole}, the $d^7$ and $d^8L$ configurations are nearly degenerate, being separated by only about 0.2~eV, which is considerably less than the hopping integrals  mixing them. In particular, the charge transfer energy from $d^7$ to $d^8L$ is $\epsilon_p-U/2$, which is distinct from the energy cost from $d^8$ to $d^9L$, which is $\epsilon_p+U/2$. 

In fact, as discussed below, turning on various hybridizations results in GS configurations involving $d^7$ states that carry substantial weights in both ambient-pressure and 15.3~GPa systems. Their weights are about half that of the dominant configuration, indicating the essential role of $d^7$ states in low-energy physics. 
The observation of a substantial $d^7$ involvement is consistent with a picture of mixed-valence fluctuating ground state (GS), namely that the nominal Ni valence of $+2.67$ arises from a mixture of Ni$^{2+}$ and Ni$^{3+}$ states, although the $d^8L$ contribution is considerably stronger than the $d^7$ contribution. 

\section{Results and discussion}
\label{sec:results}

\begin{figure*}
\psfig{figure=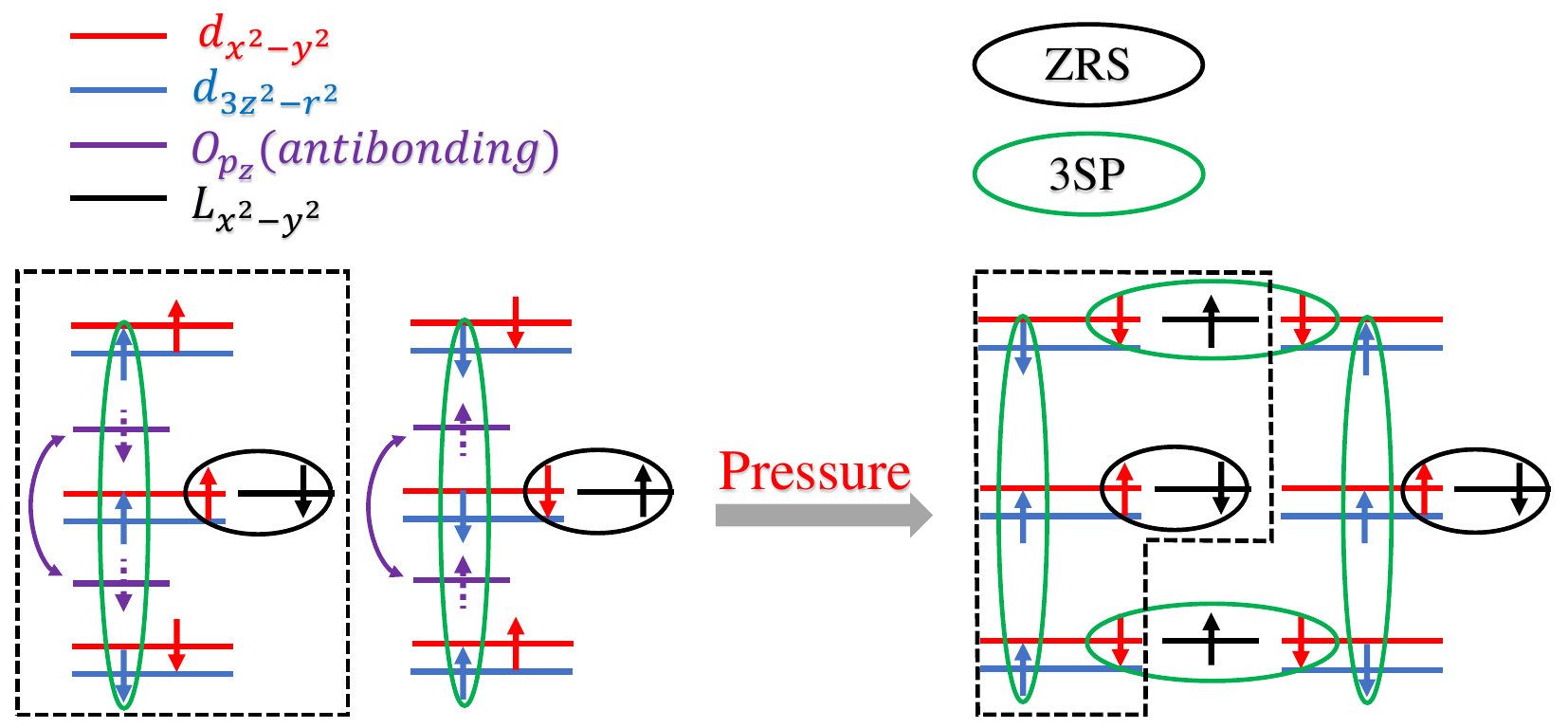, width=0.9\textwidth, clip}
\caption{\jgw{Speculated possible configurations} of the two neighboring 8-hole clusters illustrating the pressure induced redistribution of the oxygen hole states. \jgw{The configurations inside the black dashed box correspond to the dominant hole distributions obtained from numerical calculations (as shown in Fig.~\ref{8_wgt}), while the neighboring clusters are inferred schematically based on these results.} 
At ambient pressure, apart from the Zhang-Rice singlet (ZRS) in the central layer, the other hole resides on the antibonding combination of two interlayer apical oxygens and thereby couples with the 3-spin-polaron (3SP) formed by three $d_{z^2}$ orbitals.
At high pressure, this hole reverts to the outer layers and forms an in-plane 3SP with the neighboring cluster.} 
\label{cartoon}
\end{figure*}

Our major result is illustrated in Fig.~\ref{cartoon}, which shows the pressure induced redistribution of the oxygen hole states. Notably, at ambient pressure, starting with all three Ni being nominally 2+ valence, one of the extra two holes is preferentially confined in the central layer's in-plane O 2p orbitals coupling with the $x$ orbital in a Zhang-Rice singlet fashion similar to the cuprate superconductors; while the other hole resides on the antibonding combination of two interlayer apical oxygens and thereby couples with the 3-spin-polaron (3SP)~\cite{3sp1,3sp2,3sp3} formed by three $z$ ($d_{z^2}$) orbitals, namely a state expressed as 
$\sqrt{\frac{1}{6}}(\ket{d_{z^2}^{\uparrow}d_{z^2}^{\uparrow}d_{z^2}^{\downarrow}}+\ket{d_{z^2}^{\downarrow}d_{z^2}^{\uparrow}d_{z^2}^{\uparrow}}) - \sqrt{\frac{2}{3}}\ket{d_{z^2}^{\uparrow}d_{z^2}^{\downarrow}d_{z^2}^{\uparrow}}$. In this way, the remaining $x$ ($d_{x^2}$) orbitals of two outer layers are decoupled from the 3SP of $z$ orbitals because the strong $d^7$ admixture discussed below would suppress the Hund's rule coupling.  

In addition, the left side of Fig.~\ref{cartoon} also reveals the in-plane spin orientation alternation carried by the two outer layers with interlayer antiferromagnetic correlation. 
However, at high pressure, the hole in the apical $p_z$ orbital involving all three $z$ orbitals reverts to locating in $x^2-y^2$ like orbitals of the outer layers, which is depicted in Fig.~\ref{cartoon} as forming a 3SP in the outer layers with neighboring cluster. \jgw{Additionally, another independent study reports a similar cross-layer trimer\cite{jiang2025pressure}.} Hence, the net result is one hole per two in-plane oxygen $p_{x/y}$ orbitals of outer layers as in the case of La$_3$Ni$_2$O$_7$ but now in addition to one hole in $x^2-y^2$ symmetry per Ni in the central layer. Concurrently, three $z$ orbitals form a 3SP where the central $z$ spin is antiferromagnetically coupled with each of the outer layer $z$ spins via superexchange. This out-of-plane 3SP has different nature from the conventional in-plane 3SP involving O-$2p$ and Ni-$x$ that has much larger antiferromagnetic exchange~\cite{3sp3}.

\begin{figure*}[t]
\psfig{figure=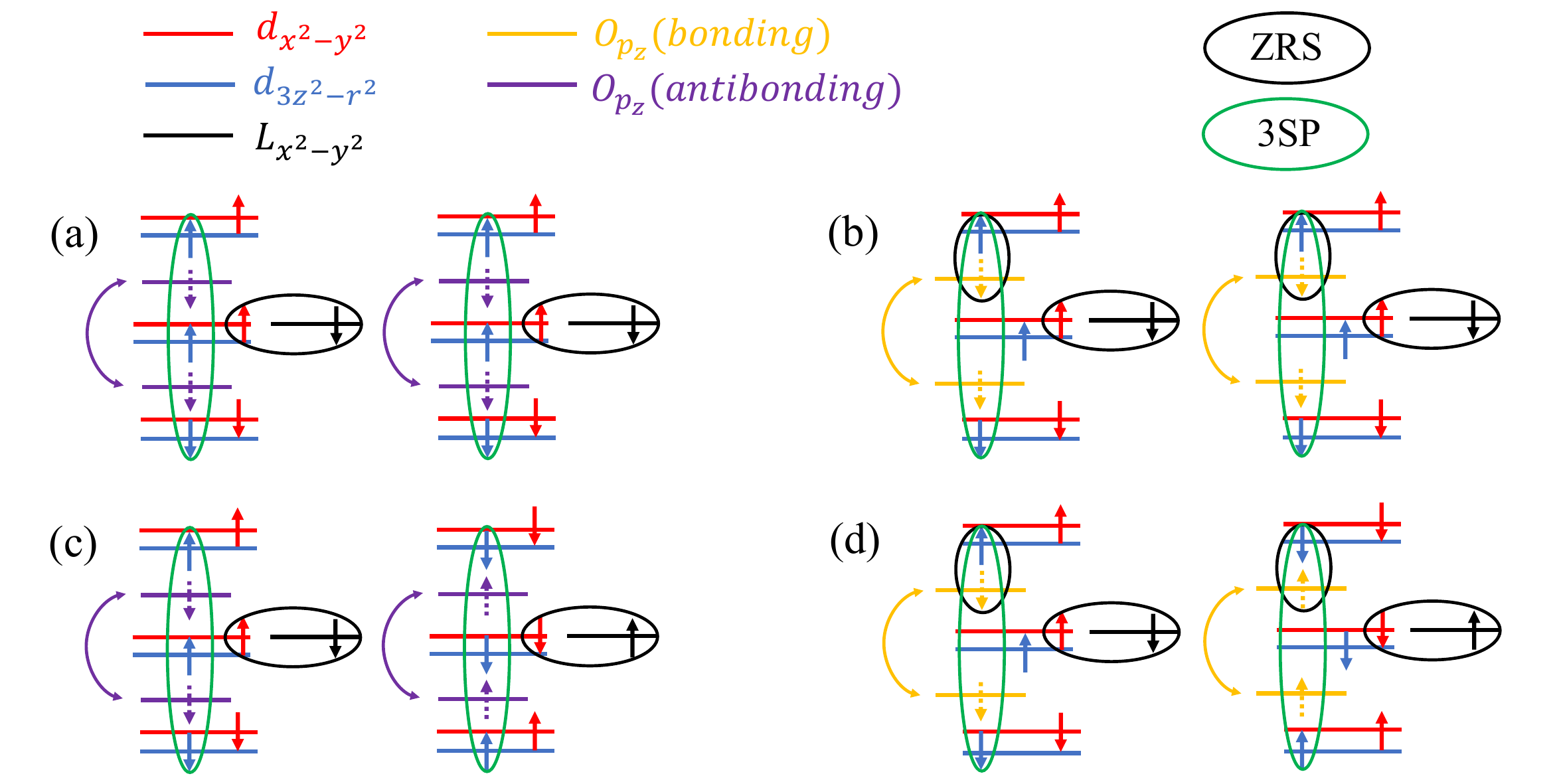, width=0.9\textwidth, trim=0.3 0 0.3 0, clip}
\caption{\jgw{Speculated possible configurations} of two neighboring clusters at ambient pressure with (a-b) spin parallel and (c-d) spin opposite between clusters. The right panels (b,d) show the bonding combination case of two interlayer apical O in spite of its higher energy than the antibonding combinations of left panels (a,c).} 
\label{cartoon0P}
\end{figure*}

\jgw{We remark that the physical picture in Fig.~\ref{cartoon}, as well as those in Fig.~\ref{cartoon0P} and Fig.~\ref{79cartoon} discussed later, are based on our concrete calculation of a single isolated cluster combined with reasonable speculated extension to two neighboring clusters. A more involved calculation explicitly considering the inter-cluster interaction is formidable and out of the scope of the current research.}

The nature of GS is governed by the competition among the Hund's coupling, the $x$-$p$ exchange through $t_{pd}$, and the $z$-$p_z$ exchange via $t_{dO}$. In addition, we have the interlayer $z$-$z$ superexchange that is active in the case of no holes in the intervening $p_z$ orbital.  
In what follows, we will provide more details of our numerical calculations of both the ambient and 15.3 GPa pressure. \jgw{Given that the $t_{dO}$ and $t_{pd}$ hybridization are two most significant parameters, for each pressure, by fixing other parameters to be their DFT derived values, our investigation spans $t_{pd}$ ranging from 0.9 to 1.1 times its DFT value with the ratio $t_{dO}/t_{pd}$ fixed at the DFT value, which is highlighted as a red star in each figure. Besides, the spin-orbit coupling is neglected for simplicity, which should not drastically modify our major results~\cite{zhang2024prediction,Dissecting,collectivespin}}.

\subsection{GS weight distribution: 8-hole system}
\begin{figure}[t!]
\psfig{figure=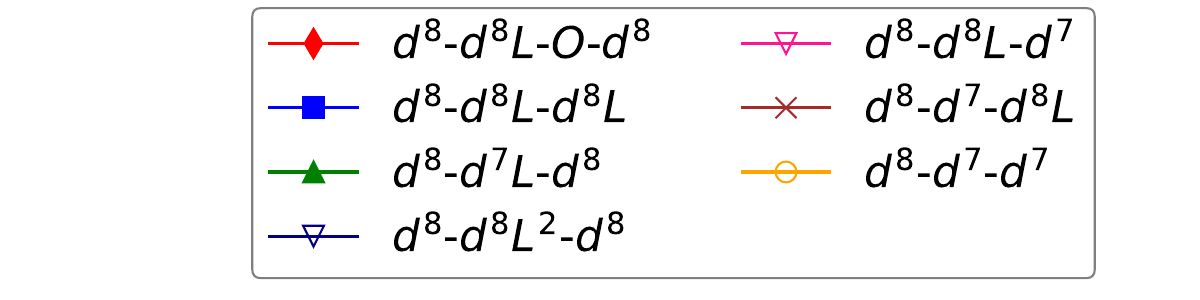, width=.45\textwidth, clip}
\psfig{figure=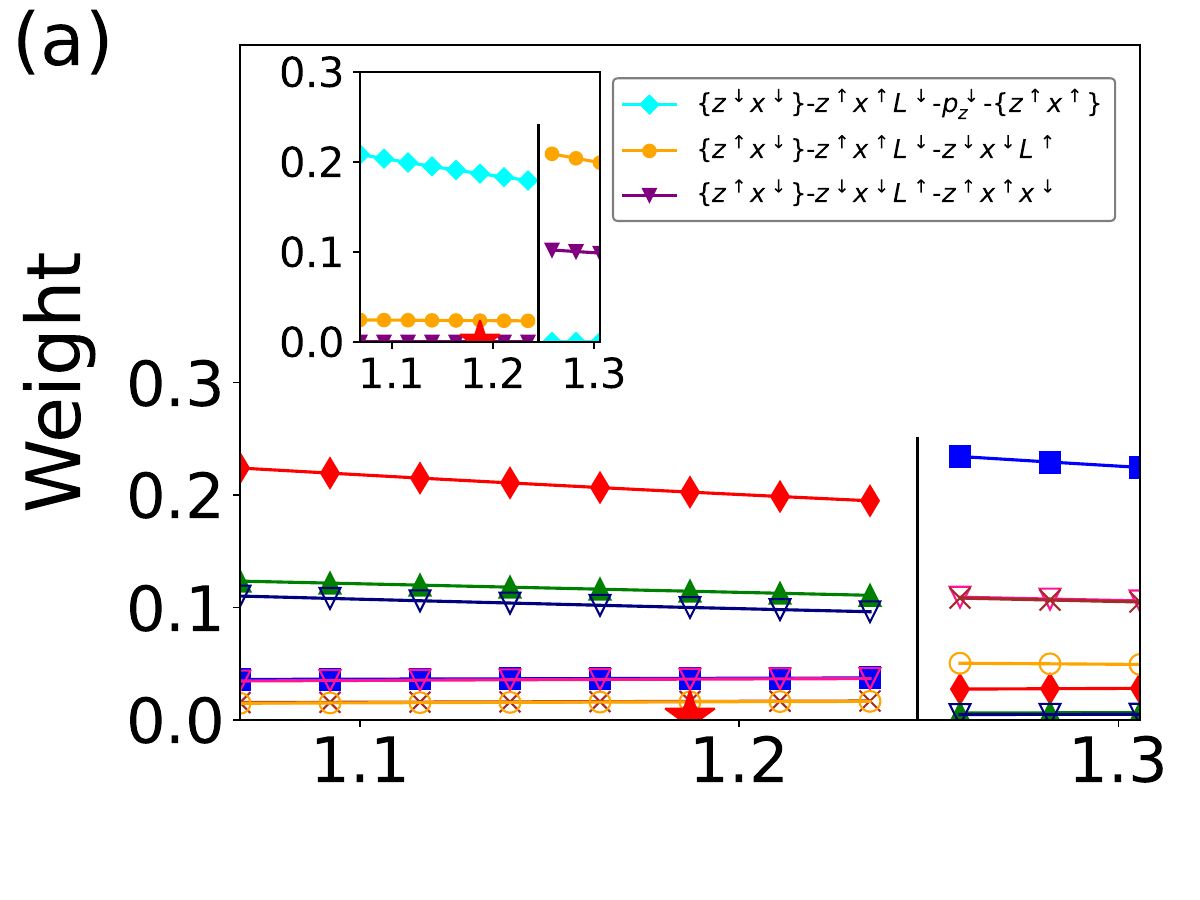, width=.45\textwidth, trim=0 40 0 0, clip}
\psfig{figure=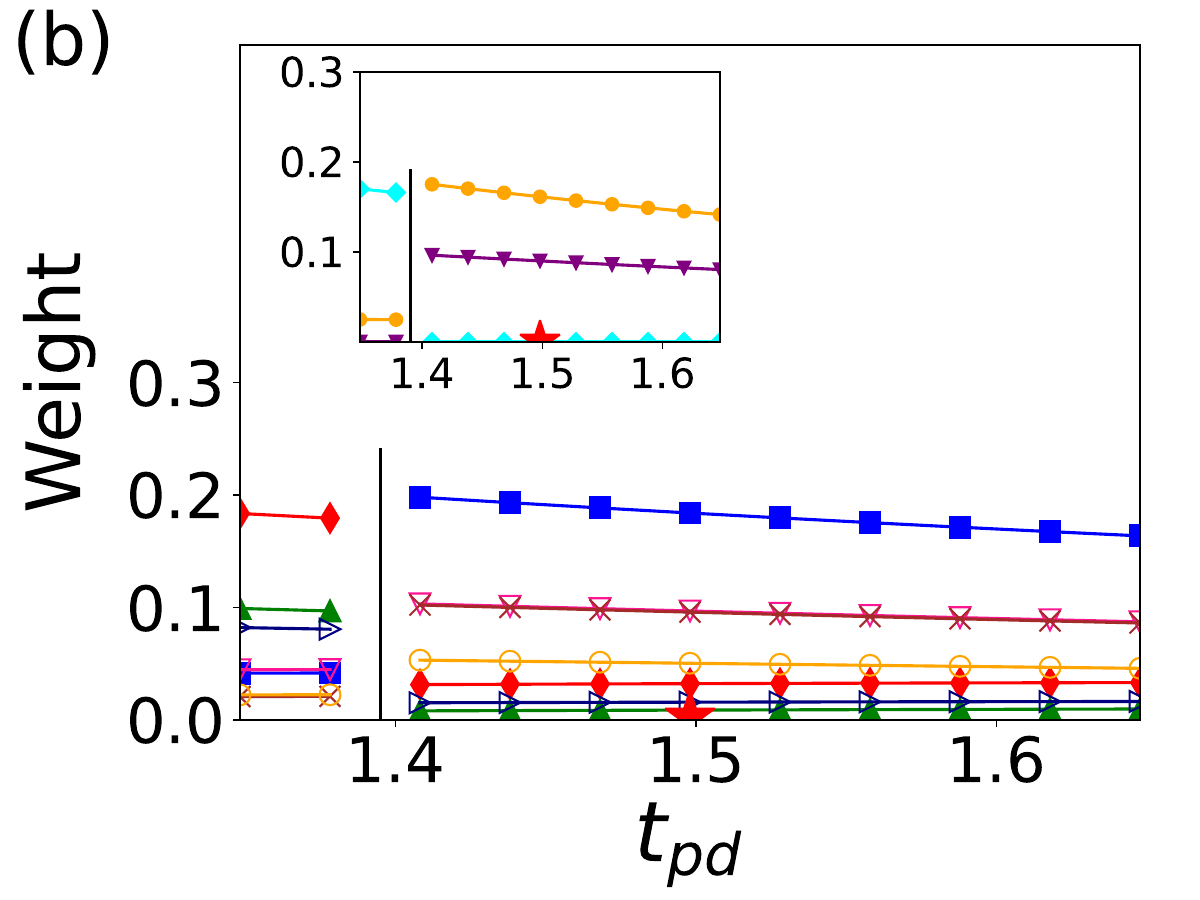, width=.45\textwidth, clip}
\caption{Comparison of GS weight distributions with $t_{pd}$ at a fixed ratio $t_{pd}/t_{dO}$ between (a) ambient and (b) 15.3 GPa pressure. The red star denotes the DFT parameters. The \{.\} symbol denotes two hole spins forming triplet states. The inset provides more details on the leading configurations. The $d_{3z^2-r^2}$ and $d_{x^2-y^2}$ orbitals are labeled as $z$ and $x$ for simplicity. }
\label{8_wgt}
\end{figure}

Fig.~\ref{8_wgt} compares the weight distribution with $t_{pd}$ of the GS between 0 GPa and 15.3 GPa, which exhibits qualitatively distinct features. At ambient pressure (red star of panel a), the dominant configuration is $d^8$-$d^8L$-$O$-$d^8$, where each layer hosts one $d^8$ state and the two remaining holes reside on the in-plane and apical oxygen, respectively. The $L$ hole on in-plane oxygen has $x^2-y^2$ symmetry to form a Zhang-Rice singlet (ZRS) (well known in hole-doped cuprates) with the Ni-$x$ orbital. 
As already shown in the left panel of Fig.~\ref{cartoon}, the interlayer $z$-$z$-$z$ 3SP hybridizes with the antibonding combination of two interlayer apical $p_z$ orbitals. We have determined that the total spin of this out-of-plane 3SP is 1/2. The strong hybridization with the antibonding O-$p_z$ orbital results in a large exchange interaction that decouples the $z$ spins from the $x$ orbital on the same Ni, i.e. the Hund's rule coupling is strongly suppressed because of the ``dilution'' effect. 

\begin{figure}[t]
\psfig{figure=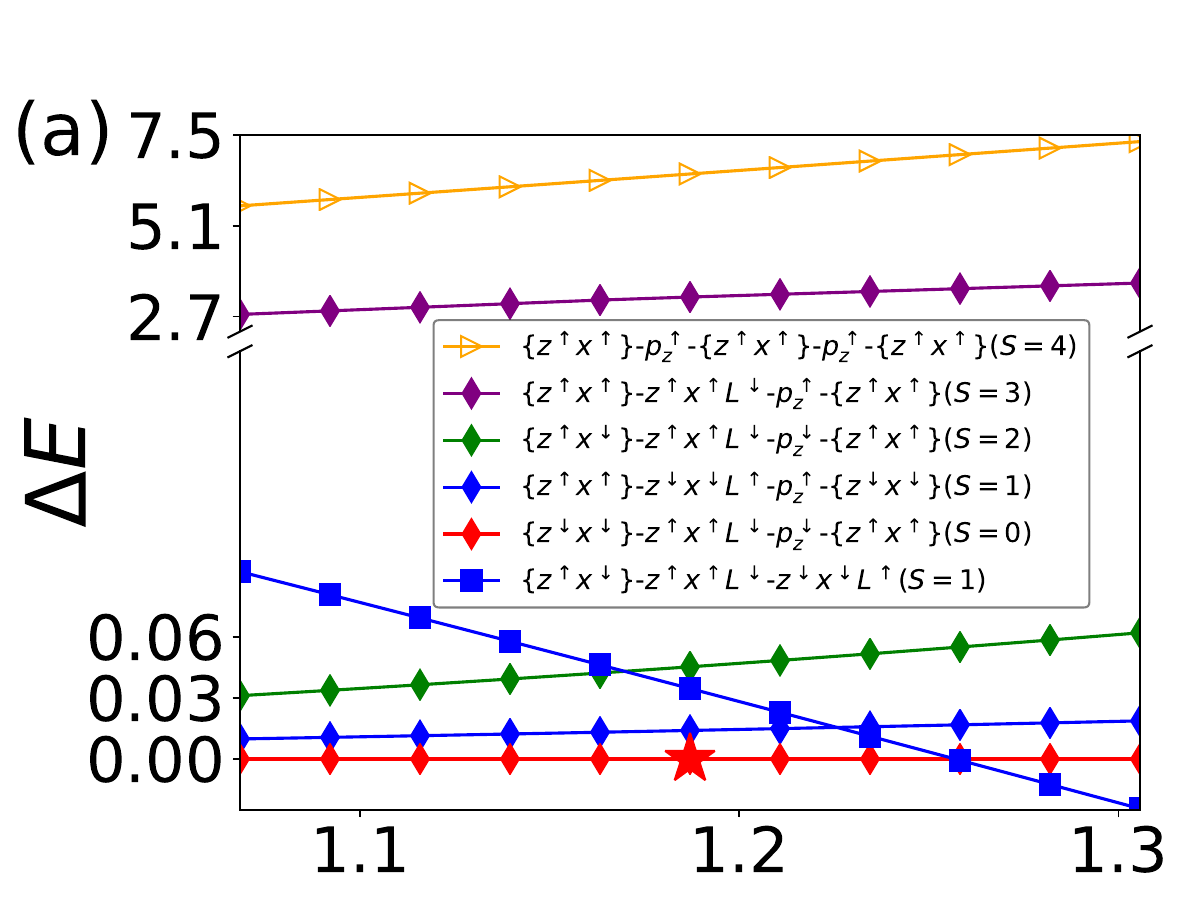, width=.45\textwidth, clip}
\psfig{figure=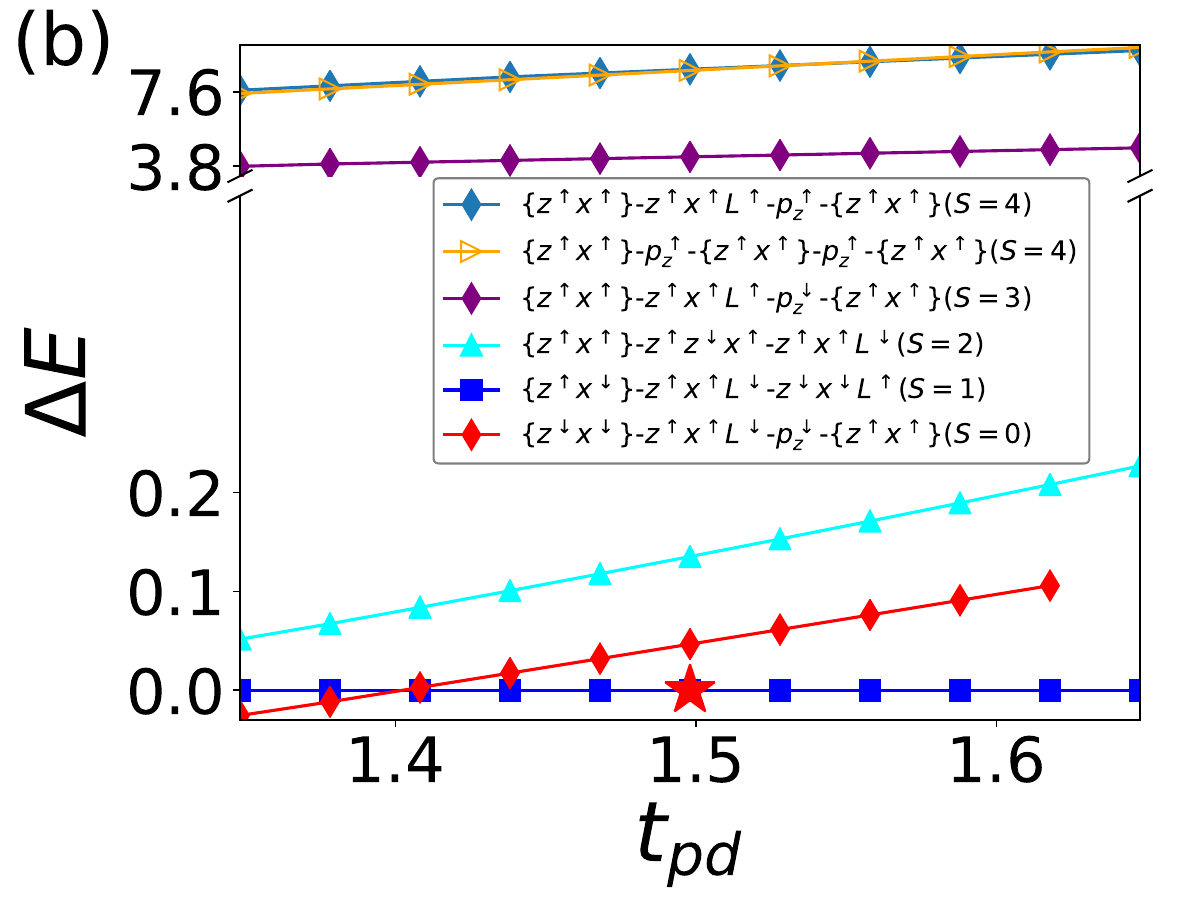, width=.45\textwidth, clip}
\caption{Excited energy splittings between spin multiplets relative to the ground state as a function of $t_{pd}$ at fixed $t_{pd}/t_{dO}$ for (a) ambient and (b) 15.3 GPa pressure. The red star denotes the DFT parameters. Labels indicate the dominant configurations and their corresponding total spins. }
\label{8_val}
\end{figure}

To consider the possible ambiguity of the parameters in the realistic solid circumstances, we tuned $t_{pd}$ (with fixed ratio $t_{pd}/t_{dO}$) in Fig.~\ref{8_wgt} and uncovered that a phase transition occurs such that the dominant configuration becomes $d^8$-$d^8L$-$d^8L$. In other words, if simply treating enlarged $t_{pd}$ as a way of mimicking the high pressure, the $x$-$L_{x^2-y^2}$ hybridization becomes increasingly dominant, driving the transfer of hole from the apical O site to in-plane O sites forming two in-plane ZRS as already shown in the right panel of Fig.~\ref{cartoon}. Besides, note that these two ZRS reside on the inner and outer layer separately.
This brings us to a situation similar to that of the bilayer RP nickelates, namely that the in-plane hole in the outer layers can be in phase or out of phase since its density is one per two in-plane Ni. 

At high pressure (Fig.~\ref{8_wgt}(b)), a phase transition appears again but to the left of the red star. Combined with the transition in Fig.~\ref{8_wgt}(a), it strongly suggests the presence of a phase transition between the ambient and high pressure. This also further indicates that varying $t_{pd}$ while keeping the ratio $t_{dO}/t_{pd}$ fixed provides a reasonable way to simulate the pressure effect.

\subsection{Spin state analysis of 8-hole system}

The pressure induced phase transition between two dominant configurations revealed in Fig.~\ref{8_wgt} leads to the total spin change as well.
At ambient pressure, shown in Fig.~\ref{8_val}(a), the low-lying eigenstates with different total spins are closely spaced in energy. In particular, the energy gaps between the $S=0, 1, \text{ and } 2$ states are quite small. This arises from the dominant $d^8$-$d^8L$-$O$-$d^8$ configuration, where the $d_{x^2-y^2}$ orbital within $d^8L$ forms an in-plane ZRS with the ligand $L$ hole, leaving an unpaired spin-$\tfrac{1}{2}$ in the $z$ orbital forming 3SP with another two $z$ orbitals of the outer layers and in turn hybridizes with the antibonding $p_z$ orbital of the bridging interlayer apical O. 
As shown in the left panel of Fig.~\ref{cartoon}, this subsystem of out-of-plane 3SP together with the antibonding $p_z$ hole induces the superexchange between the two outer layers, \jgw{which can be estimated by the energy gap between different total spins in Fig.~\ref{8_val}.}

In contrast, for $S = 3$ and $S = 4$, the ZRS is broken, leading to Zhang-Rice triplet like configurations with ferromagnetic spin alignment, resulting in a substantial increase in the total energy. 
In addition, the energy splittings between $S = 0,1,2$ configurations are so small that it allows for spin state transition, as observed in the GS spin changes from $S = 0$ at ambient pressure to $S = 1$ at large $t_{pd}$, which is consistent with the phase transition around $t_{pd}\sim1.25$ in Fig.~\ref{8_wgt}(a). These observations would lead to additional phase transitions or crossovers, as observed experimentally.

\begin{figure*}
\psfig{figure=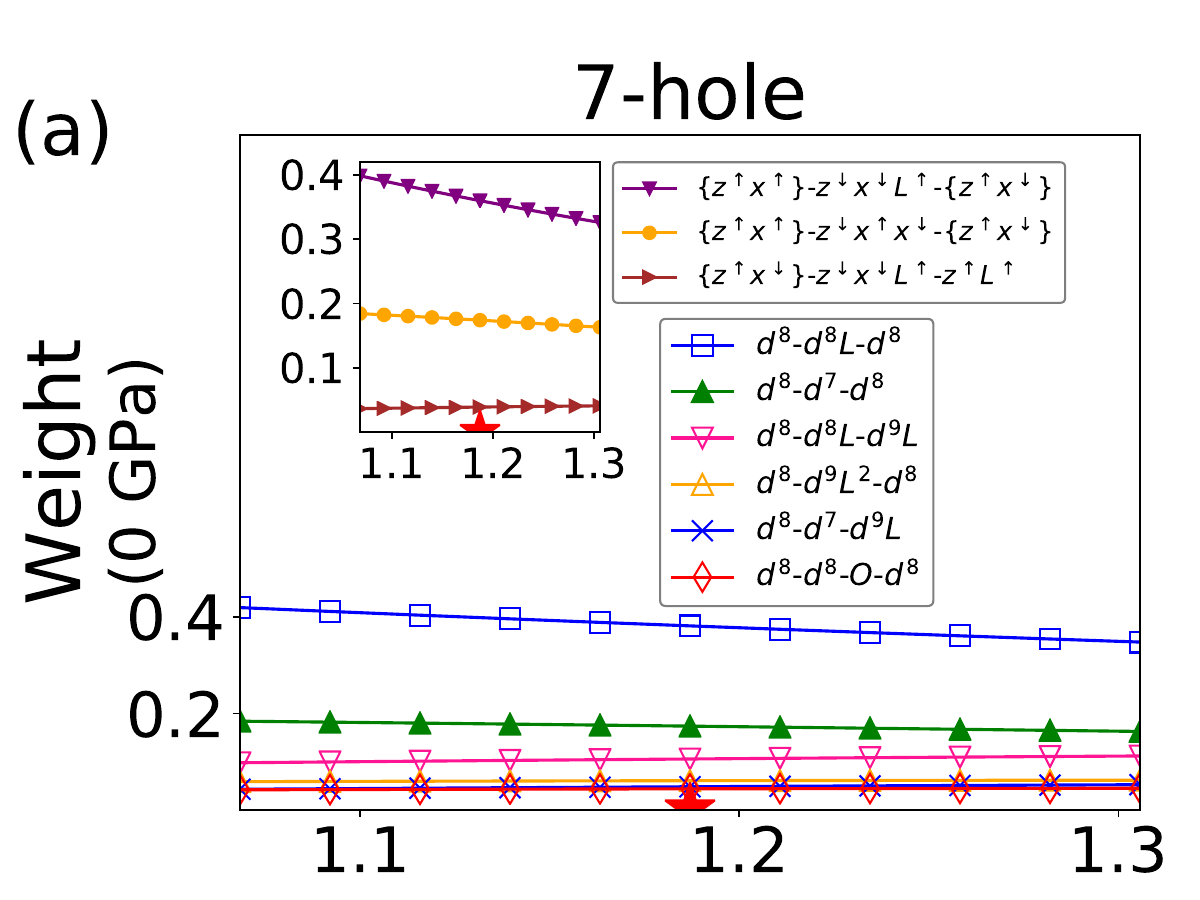,width=.45\textwidth, clip}
\psfig{figure=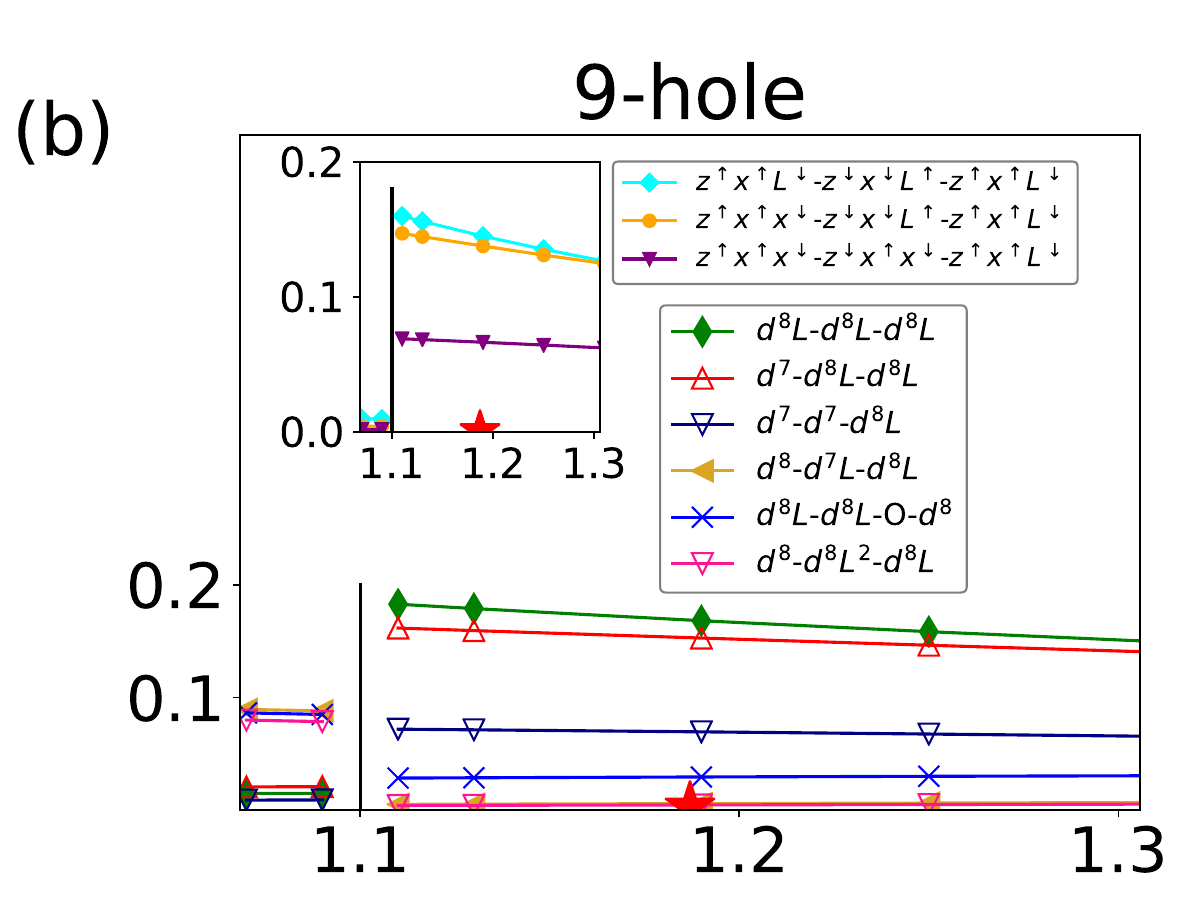,width=.45\textwidth, clip}
\psfig{figure=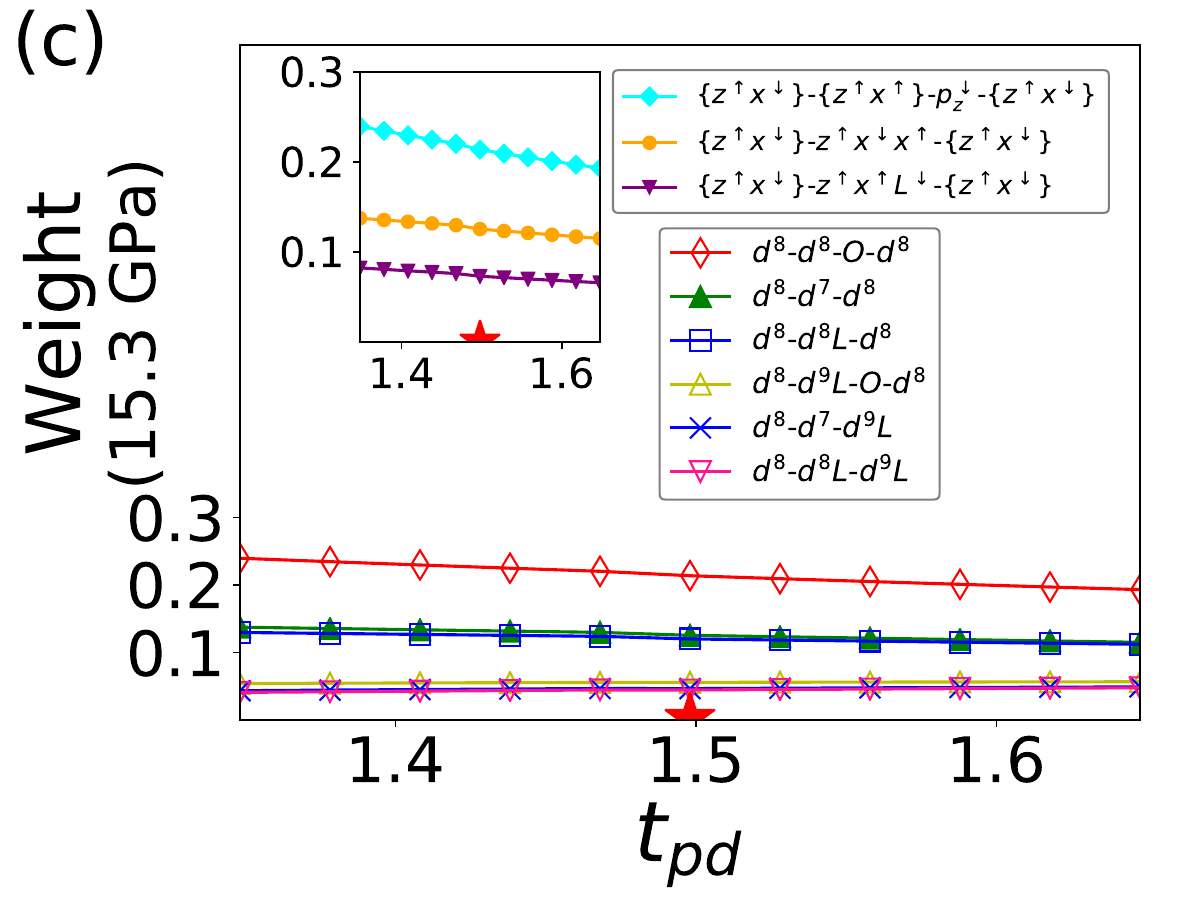,width=.45\textwidth, clip}
\psfig{figure=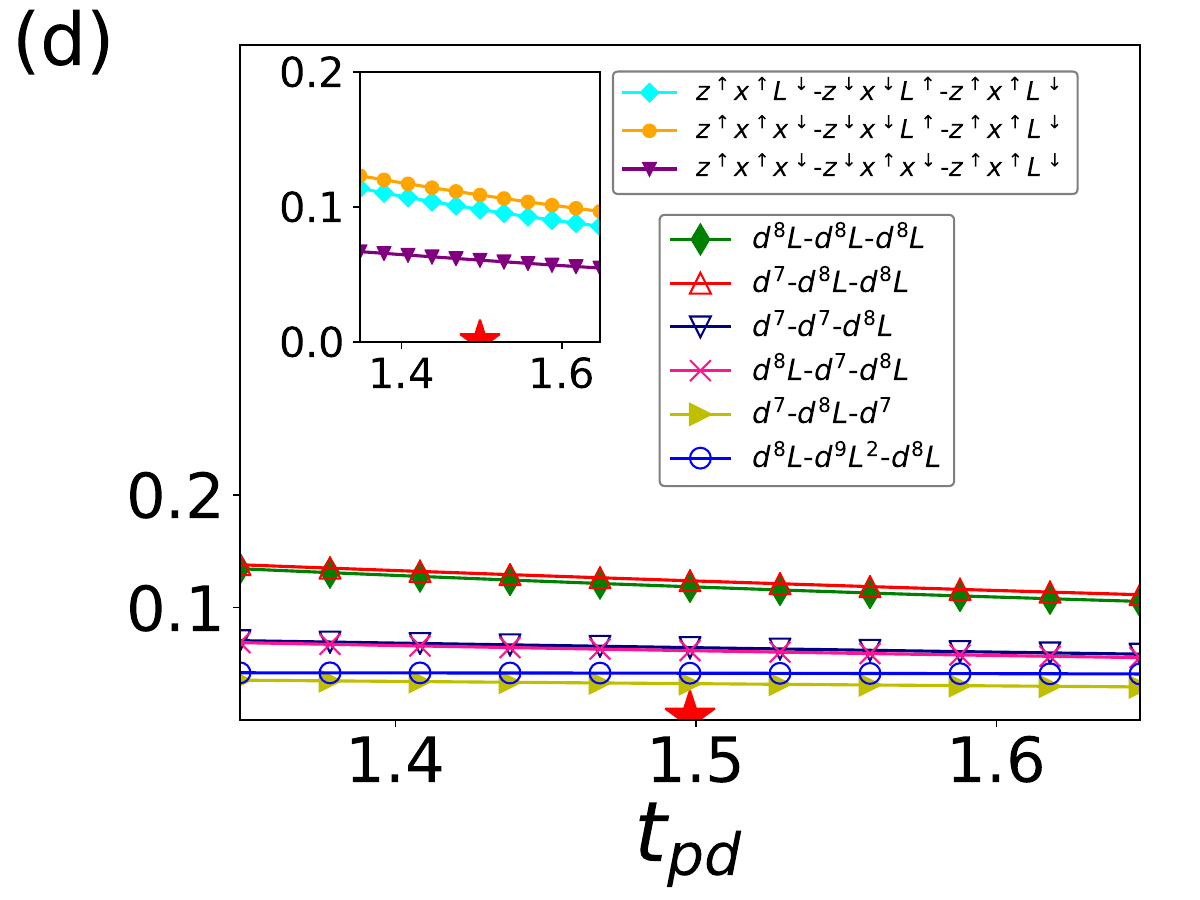,width=.45\textwidth, clip}
\caption{GS weight distribution of 7-hole and 9-hole system for both ambient and high pressure.}
\label{79hole}
\end{figure*}

Moving to Fig.~\ref{8_val}(b), clearly there is also a transition from the GS $S=0$ to $S=1$ around $t_{pd}\sim1.4$ mirroring the transition in Fig.~\ref{8_wgt}(b). \jgw{Apparently, the interlayer superexchange at high pressure is stronger than that at ambient pressure. Given that SC only emerges under high pressure for bulk La$_4$Ni$_3$O$_7$, it is plausible to expect the close connection between the interlayer superexchange and SC.
Furthermore, compared with high-pressure La$_4$Ni$_3$O$_7$~\cite{jiang2025intertwined}, La$_4$Ni$_3$O$_{10}$ exhibits both a weaker interlayer superexchange and a lower superconducting transition temperature. Besides, recent study based on the singlet triplet splitting of spin-$1/2$ $d_{3z^2-r^2}$ states in the La$_3$Ni$_2$O$_7$ compound revealed both intraplane and interplane pairing channels with the interplane pairing being stronger~\cite{Khaliulinsorbital}. Although these observation do not establish a direct connection, they strongly imply the relevance of interlayer superexchange to SC. } 

Interestingly, although the GS at 0 and 15.3~GPa differ markedly in both spin and orbital configuration, the excited states in both cases show a similar evolution, from configurations dominated by in-plane oxygen holes to those involving interlayer apical oxygens and eventually forming the $d^8$-$O$-$d^8$-$O$-$d^8$ pattern at very high-energy $S = 3, 4$ states.

Fig.~\ref{cartoon} only illustrates one reasonable configuration of two neighboring clusters. Because Fig.~\ref{8_val}(a) reveals the quite small energy difference between various configurations of different total spins, for example, the blue and red diamond curves, this strongly implies that including inter-cluster interactions, albeit out of our current simulation capability, could radically change the configurations shown in Fig.~\ref{cartoon} obtained via our isolated simulations. Therefore, more possibilities for the ambient pressure are shown in Fig.~\ref{cartoon0P}, where the upper/lower panel shows the spin parallel/opposite between clusters, respectively. \jgw{Because our calculations are limited to the isolated cluster, it is impossible to expect its agreement with the experimentally revealed spin density wave vector~\cite{dw4310,4310neutron}. In fact, the realistic situation would be a linear combination of all possible configurations together with strong quantum fluctuation between these diverse possibilities.} In addition, the panels (b) and (d) also provide the pictures of bonding combination of two interlayer apical oxygens, although their weights in GS are much smaller than their antibonding counterparts shown in panels (a) and (c). In other words, the bonding combination has much higher energy than the antibonding case focused in the present study e.g. Fig.~\ref{cartoon}. This observation emphasizes the importance of the $z$ orbital of the central layer, which plays a vital role in bridging two outer layers.





\subsection{7- and 9-hole systems}

\begin{figure*}
\psfig{figure=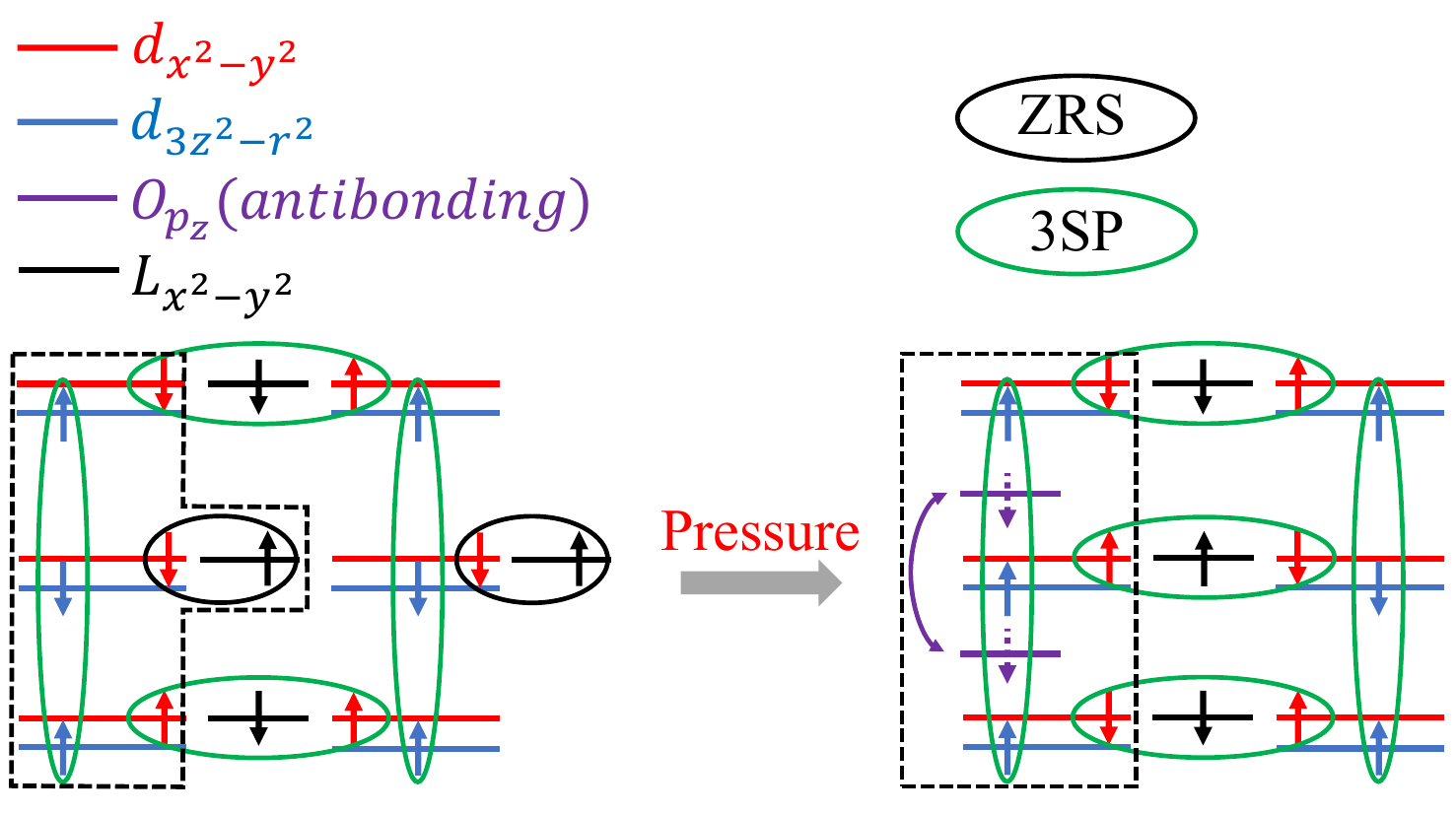, width=0.9\textwidth, clip}
\caption{\jgw{Speculated configurations of neighboring 7-hole (black dashed box, as shown in Figs.~\ref{79hole}(a,c)) and 9-hole clusters (as shown in Figs.~\ref{79hole}(b,d)). The hole distributions within each individual cluster are obtained directly from the calculations, while the two neighboring clusters are inferred from these results to illustrate the possible pressure-induced redistribution of oxygen hole states.}
}
\label{79cartoon}
\end{figure*}

In our previous investigation on bilayer La$_3$Ni$_2$O$_7$~\cite{jiang2025intertwined}, we have seen the possible role of charge density modulation between two neighboring clusters. A similar analysis can be performed for the trilayer La$_4$Ni$_3$O$_{10}$.
To this aim, we first discuss the GS weight distribution of the 7- and 9-hole systems, which are electron addition and removal states, respectively and closely related to the charge gap of the 8-hole system. Precisely, the gap is the energy difference $\Delta E = E_{7+9}-E_{8+8}$ between the GS energy of a 7- and a 9-hole cluster $E_{7+9} = E_{GS,7}+E_{GS,9}$ and that of two 8-hole clusters, $E_{8+8} = 2E_{GS,8}$.
We found that the charge gap turns out to be small, which is well within the expected bandwidth due to the inter-cluster hopping. Hence, these materials are predicted to be metals  although small gaps could occur if indeed spin and charge density waves exist, which is out of our capability to decisively determine. \jgw{Note that this rough estimation of the charge gap without the full consideration of the inter-cluster interaction differs from the metallic transport, whose evaluation is out of the scope of this work.}

Fig.~\ref{79hole} illustrates the pressure dependence of their GS weights of the dominant configurations. 
For the 7-hole system, starting from the $d^8$ states in all three layers, the high pressure induces the transition to the dominant configuration of the extra hole occupying apical O (panel c) from the in-plane O at ambient pressure (panel a).  

For the 9-hole system, unless  $t_{pd}$ is very small, the physically relevant DFT parameters at both ambient (panel b) and high pressure (panel d) result in the leading configuration of $d^8L$ in all three layers with strong involvement of $d^7$ state in one outer layer. Nonetheless, the apical O only plays a minor role in the 9-hole system. 

With the above manifestation of 7- and 9-hole system combining with the previously discussed 8-hole one, we can speculate on  more possible ordered phases suggested by these single cluster results.
Specifically, Fig.~\ref{cartoon} only illustrates the speculation of two neighboring clusters both with 8 holes. Instead, Fig.~\ref{79cartoon} reveals other speculated possibilities consisting of 7+9 holes. 
Intriguingly, the ambient pressure configuration is the same as the high pressure case of 8+8 holes, namely in-plane 3SP within two outer layers. At high pressure, all three layers possess the in-plane 3SP as well as the presence of antibonding apical O.

All proposed possible configurations of two neighboring clusters (either 7+9 holes or 8+8 holes) can be categorized by the number $N_{3SP}$ of in-plane 3SP states: (i) $N_{3SP}=0$ as shown in Fig.~\ref{cartoon}'s left side; (ii) $N_{3SP}=2$ and are located on two outer layers as shown in Fig.~\ref{79cartoon}'s left side; (iii) $N_{3SP}=3$ evenly distributed on all layers as shown in Fig.~\ref{79cartoon}'s right side. 
\jgw{We remark again that these proposed local spin and charge configurations are all consistent with the antiferromagnetism between two outer layers. Besides, they can potentially support possible in-plane SDW/CDW picture while they do not by themselves determine the experimentally observed ordered state.}


\section{Summary and outlook}
\label{sec:Summary}
In summary, the present study reveals that the undoped La$_4$Ni$_3$O$_{10}$ exhibits markedly different characteristics between ambient and high pressure. Specifically, starting with all three Ni being nominally 2+ valence, at ambient pressure, the GS is dominated by configurations with one of two extra holes preferentially confined in the central layer's in-plane O-$2p$ orbitals coupling with the $x$ orbital as a Zhang-Rice singlet; while the other hole resides on the antibonding combination of two interlayer apical oxygens and thereby couples with the 3-spin-polaron (3SP) formed by three $z$ orbitals.  

In contrast, under high pressure, the apical O contribution vanishes and two additional holes both reside in two neighboring NiO$_2$ layers. In particular, one hole resides on the in-plane oxygen of the outer layers and forms an in-plane 3SP with the neighboring cluster. Therefore, the net result is one hole per two in-plane O of outer layers as in the case of La$_3$Ni$_2$O$_7$ but now in addition to one hole in $x^2-y^2$ symmetry per Ni in the central layer. Concurrently, similar to the ambient pressure situation, three $z$ orbitals form a 3SP where the central $z$ spin is antiferromagnetically coupled with each of the outer layer $z$ spins via superexchange. This out-of-plane 3SP has a different nature from the conventional in-plane 3SP involving O-$2p$ and Ni-$x$ that has much larger antiferromagnetic exchange~\cite{3sp3}.

\jgw{Based on these concrete calculation of an isolated cluster, for the situation of two neighboring clusters, we speculated various possible in-plane spin orientation alternation carried by the two outer layers with interlayer antiferromagnetic correlation.} 
\jgw{In our proposed picture, the pressure transfers holes from apical to in-plane oxygen orbitals of outer layers, generating in-plane 3SP-like quasiparticles that act as mobile carriers coupled by interlayer superexchange; while interplane 3SP-like states may provide the pairing glue. Since the low-pressure phase lacks freely propagating in-plane quasiparticles, this scenario naturally favors SC in the high-pressure phase.}
A more involved calculation explicitly considering the inter-cluster interactions, in spite of its formidable complexity, is strongly desired in future research to uncover more distinct physics in this nickelate superconductor.

\section{Acknowledgments}
Guiwen Jiang and Mi Jiang acknowledge the support of the National Natural Science Foundation of China (Grant No.12174278) and State Key Laboratory of Surface Physics and Department of Physics in Fudan University (Grant No.KF2025\_12). Mi Jiang also acknowledges the support of the China Scholarship Council (CSC) and the hospitality of Karsten Held at TU Wien. Liang Si acknowledges support from the National Natural Science Foundation of China (Grant No.12422407). George A. Sawatzky is supported by the Quantum Matter Institute (QMI) at the University of British Columbia and the Natural Sciences and Engineering Research Council of Canada (NSERC). The computational work was carried out on the resources provided by Soochow University and the National Supercomputing Center in Xi’an, hosted by Northwest University.

\bibliography{main}

\end{document}